# Giant staggered Dzyaloshinskii-Moriya vectors emerged in synthetic antiferromagnets with spatially alternating stress

Yuuga Iwata[1], Kensuke Hayashi[1], Reiju Furumaki[1], Satoshi Iihama[1], Masahiro Sato[2], and Takahiro Moriyama[1,3†]

[1]*Department of Materials Physics, Nagoya University, Nagoya, Aichi 464-8603, Japan*

[2]*Department of Physics, Chiba University, Chiba, Chiba 263-8522, Japan*

[3]*Research Center for Crystalline Materials Engineering, Nagoya University, Nagoya, Aichi 464-8603, Japan*

## Abstract

Dzyaloshinskii-Moriya (DM) interaction is a source of chiral magnetic physics, and it manifests microstructural symmetry which could link to exotic magnetic and electronic properties such as altermagnetism. For a new paradigm in magnetism and spintronics, it is crucial to have control on the DM interaction as well as ordering of the DM vectors which characterize the DM interaction. Here, we report on a strong interlayer DM interaction emerged in Co/Ru magnetic superlattices whose structural symmetry is unambiguously broken by alternating mechanical strains developed using the peculiar technique of our stress mechanism. Moreover, theoretical analyses reveal that our magnetic superlattices host a staggered order of the DM vectors. Our results indicate that the spatially varying strains with our breakthrough technique are quite effective in breaking the symmetry of the system and prove that it can control the DM interaction strength as well as the DM vector order.

[†] corresponding to moriyama.takahiro.a4@f.mail.nagoya-u.ac.jp

Symmetry is one of the fundamental concepts in physics. Microscopic structural symmetry often manifests physical properties in solids. In magnetism, Dzyaloshinskii-Moriya (DM) interaction [1,2] is one of the important consequences of such structural symmetry, which link to varieties of rich chiral magnetic orders such as skyrmions [3], chiral magnetic domain walls [4], and non-colinear frustrated magnetic structures [5].

DM interaction is an antisymmetric exchange interaction where neighboring spins $\mathbf{S}_i$ and $\mathbf{S}_j$ align based on the energy $E_{\mathrm{DM}} = -\mathbf{D} \cdot (\mathbf{S}_i \times \mathbf{S}_j)$, where $\mathbf{D}$ is the Dzyaloshinskii-Moriya (DM) vector, the direction of which is determined by Moriya rules which follow the microstructural symmetry of the system [6]. For the magnitude of the DM vector to be non-zero, there should be spin orbit coupling in addition to the symmetry breaking in the system. DM interaction induces a clockwise or counter-clockwise rotation between neighboring spins depending upon the polarity of the DM vector, while a symmetric exchange interaction, or the Heisenberg exchange interaction, whose energy described by $E_{\mathrm{Heis}} = -J(\mathbf{S}_i \cdot \mathbf{S}_j)$ with the exchange constant $J$ favors parallel or antiparallel alignment of the neighboring spins depending upon the sign of $J$. The Heisenberg exchange interaction is a main source of the ferro- and antiferro-magnetism while the concept of DM interaction was first conceived for explaining the weak magnetism in hematite ($\alpha$-$Fe_2O_3$) having canted spin alignments between magnetic ions $Fe^{3+}$[1]. The DM interaction and ordering of the DM vectors in solids have recently revisited amid surge of the concept of altermagnets [7,8] which newly categorizes magnetic materials having particular crystalline symmetries but previously thought to be antiferromagnets or weak ferromagnets including hematite mentioned above [7].

The significance of DM interaction has been extended to magnetic heterostructures where the heterointerfaces are the source of a broken inversion symmetry.

An interface between a ferromagnetic layer and a heavy metal layer, such as Pt, Pd and Ir, leads to the so-called *interfacial* DM interaction acting between the spins within the ferromagnetic layer [9]. This is termed against the “bulk” DM interaction seen in ionic crystals like hematite. This interfacial DM interaction together with the spin orbit-toque (SOT) has attracted intense attention for spintronic applications, such as current-driven skyrmion devices, as both the interfacial DM interaction and SOT can be controlled to some extent by interfacial engineering and material combinations [10].

The exchange interactions are subject not only between spins but also between magnetic layers. In the magnetic superlattices, such as $[Co/Cu]_x$ and $[Fe/Cr]_x$, a strong interlayer exchange interactions between magnetic layers across a non-magnetic metallic layer is often observed [11,12], which is explained by the Ruderman–Kittel–Kasuya–Yoshida (RKKY) interaction [13] associated with exchange of the iterant electrons through the non-magnetic layer. The RKKY interaction is a Heisenberg-type interaction whose energy is described by $E_{\mathrm{RKKY}} = -J_{\mathrm{RKKY}}\left(\mathbf{M}_j \cdot \mathbf{M}_{j+1}\right)$, where $\mathbf{M}_j$ is the macroscopic spin in the *j*-th magnetic layer separated by a non-magnetic layer (note we assume spins are uniformly aligned in each magnetic layer.). The magnitude and the sign of $J_{\mathrm{RKKY}}$ varies with the thickness of the spacer layer [11,13].

Very recent investigations have shown an appreciable DM interaction acting between magnetic layers across a non-magnetic spacer layer [14,15] in addition to the RKKY interaction. This *interlayer* DM interaction, $E_{\mathrm{IL-DM}} = -\mathbf{D}_{\mathrm{IL-DM}} \cdot \left(\mathbf{M}_j \times \mathbf{M}_{j+1}\right)$, originates from the broken inversion symmetry at the multilayer interfaces and a strong spin orbit interaction provided by the heavy metals in the spacer layer. The studies have reported the interlayer DM interaction energy of as large as $E_{\mathrm{IL-DM}} \sim 0.01$ mJ/m$^2$ which is significant enough to create chiral spin textures. The potential of the interlayer DM interaction has already been demonstrated in spintronic memory applications where the chiral spin texture helps the field-free current induced magnetization switching [16,17].

DM interaction is, therefore, a source of chiral magnetic physics and it manifests microstructural symmetry which could link to exotic magnetic and electronic properties such as altermagnetism. It is undoubtedly important to have control of the DM vector for a new paradigm in magnetism and spintronics. Although the interfacial DM interaction seems to be more controllable compared to the bulk DM interaction relying on the crystalline symmetry, it is still the interfacial effect that is inevitably diluted by the volume of the system, and it provides no control of the DM vectors as a collective order parameter. It is therefore desirable to seek different principles that can induce DM interaction which is scalable with respect to the volume of the system.

Here, we report on a strong interlayer DM interaction emerged in Co/Ru magnetic superlattices whose structural symmetry is unambiguously broken by mechanical alternating strains in a scalable manner. By means of the state-of-the-art biaxial stress mechanism in an ultra-high vacuum deposition system, mechanical strains in different directions are induced in each magnetic layer so that the rotational symmetry around the film normal is broken in the entire superlattice. Such magnetic superlattices are found to have a staggered order of the DM vectors with a remarkably large magnitude. We emphasize that the present study essentially develops a novel way of artificially controlling the DM interactions as well as their ordering in the magnetic superlattices.

Mechanical strains modify electrical and magnetic properties via the deformation of the crystalline lattice. In particular, magnetism and strain are coupled through the magnetoelastic effect. In the context of flexible spintronics [18], strains of a few percents can be applied in the magnetic layers deposited on a flexible substrate and change the magnetic anisotropy energy by ~ 50 kJ/m$^3$ [19]. Most of such studies are, however, on the system with a spatially uniform strain which is insufficient to create the symmetry breaking for DM interaction to emerge.

Figure 1 illustrates the film deposition technique with the alternating stress mechanism and a $[Co/Ru]_x$ superlattice with the spatially alternating compressive strains. We employ a 0.1 mm-thick polyethylene naphthalate (PEN) sheet as a flexible substrate. The substrate was cut into 20 x 20 mm$^2$ square and set into a home-made stress mechanism designed to apply a compressive force that smoothly bends the substrate in two orthogonal directions (See more detailed descriptions in the Supplementary Materials [20]).

First, a compressive force $\mathbf{F}_1$ bends the substrate as shown in Fig. 1(a). The Ru and Co layers are then subsequently deposited. When the compressive force is released and the substrate elastically recovers its flat surface, a compressive residual stress $\mathbf{N}_1$ is built in the layers along the direction in which $\mathbf{F}_1$ was applied. Next, before depositing another set of Ru and Co layers, a compressive force $\mathbf{F}_2$ is applied to bend the substrate in the orthogonal direction to the first bend. The second set of the Ru and Co layers then undergoes the compressive residual stress $\mathbf{N}_2$ orthogonal to $\mathbf{N}_1$. The procedure is repeated without breaking the vacuum. We then obtain the magnetic superlattice shown in Fig. 1 (b) with spatially alternating $\mathbf{N}_1$ and $\mathbf{N}_2$ with $\mathbf{N}_1 \perp \mathbf{N}_2$ and $|\mathbf{N}_1| = |\mathbf{N}_2|$. This superlattice structure with the spatially alternating compressive stress is found to possess 2-fold rotational symmetry with a rotational axis along the film normal ($z$-axis) and a translational symmetry along $z$-axis.

The entire layer stack is designed to be PEN sub./Ru 2.0 nm/[Co 2.0 nm/ Ru $t_{Ru}$ nm]$_9$ /Co 2.0 nm/ Ru 5.0 nm (referred to as $[Co/Ru]_{10}$ superlattice hereinbelow), where 10 of the Co layers are separated by the Ru spacer layers with the 2.0 nm buffer layer and the 5.0 nm capping layer of Ru. The Ru and Co layers were deposited by d.c. magnetron sputtering with the base pressure of $10^{-6}$ Pa. From the bend curvature of the substrate with $\mathbf{F}_1$ and $\mathbf{F}_2$, the compressive strain of ~0.8% is estimated to be retained in each layer.

Hereafter, we focus on the $[Co/Ru]_{10}$ superlattice having antiferromagnetic coupling between Co layers with $t_{Ru}$ = 0.5, 0.8, and 1.0 nm. The X-ray reflectivity measurements in Fig. 1(c) clearly show the superlattice peaks reflecting the Co/Ru repetitions, indicating the layer structures are well maintained on the flexible substrate. The repetition periods are found to be 2.7, 2.5, and 2.2 nm for $t_{Ru}$ = 1.0, 0.8, and 0.5 nm, respectively. The systematic difference of 0.3 nm between the estimated period and the designed [Co/Ru] thickness ($t_p = t_{Ru} + 2.0$ nm) might be due to material intermixing at the interfaces or small errors in our calibration of the sputtering rate.

Figure 2 compares typical magnetization curves measured with the field along $x$ direction for the three different layer stacks with and without alternating residual stress. We note that there are no significant differences in the hysteresis curves measured along $x$- and $y$-directions. The data are therefore displayed only with $x$-direction. Without presence of the stress, all the magnetization curves show a negligibly small remanence at zero field and a high saturation field, indicating the neighboring Co layers are antiferromagnetically coupled at zero field. The saturation field increases from 2 to 6 T with decreasing $t_{Ru}$, which corresponds to the RKKY interaction energy of $E_{\mathrm{RKKY}} = -3.0 \sim -1.0$ [mJ/m$^2$]. The observed saturation magnetization, calculated based on the total magnetic moment divided by the volume of the total Co layers, was found to be about 20 % smaller than the typical saturation magnetization of Co. The reduction of the magnetization could be attributed to a magnetic dead layer at the Co/Ru interfaces [21,22].

Remarkable differences in the magnetization curves are seen when they are compared with the samples with the alternating stress. The shape of the magnetization curves drastically changes with a few notable characteristics, and they are more emphasized in thinner $t_{\mathrm{Ru}}$ samples. One is that the samples with the alternating stress show a more rapid increase of magnetization in the small field ranging from 0 to 2 T and show a non-zero remanence (See the arrow in Fig. 1 (a)). Another is that the ones with the alternating stress show a larger saturation field than the ones without (See in particular Figs. 2 (a) and (b)).

The last characteristic denotes that the magnetic moment of the Co layers is almost invariant regardless of whether the strains are applied or not. This essentially indicates that thickness of Co layers is invariant and perhaps the same for the Ru layers, implying that the stress mechanism employed in the deposition system does not affect the film deposition other than the introduction of stress. On the other hand, the first two characteristics pointed above essentially infer that the antiferromagnetic coupling becomes weaker in the low field, and it becomes stronger in the higher field, which do not seem to be explained by either the stress modification of the RKKY interaction or that of the magnetic anisotropy [19], as discussed next.

We then investigate the effect of the stress on the magnetic anisotropy of the Co layer and how it impacts the $[\mathrm{Co/Ru}]_{10}$ superlattice with antiferromagnetic coupling. Figure 3 (a) shows the magnetization curves of Ru 2.0 nm/ Co 2.0 nm/ Ru 5.0 nm with a uniform residual stress on the entire trilayer with $\mathbf{N}_1$ in a given direction. The magnetization curves measured with the external field along and perpendicular to $\mathbf{N}_1$ differ due to the stress-induced magnetic anisotropy. The induced magnetic anisotropy energy is estimated, by comparing the magnetization curves, to be as large as $K_s = 2.7$ [kJ/m$^3$] which is equivalent to ~5 mT of the magnetic anisotropy field along the direction of the strain. In the same way, the [Co 2.0 nm/ Ru 0.8 nm]$_{10}$ superlattices with a uniform residual stress (the applied strains are in the same direction in all the layers) were characterized as shown in Fig. 3 (b). No differences in the magnetization curves indicate that the stress-induced magnetic anisotropy has negligible effects on the magnetization process in the superlattice. This is a quite reasonable consequence as the induced magnetic anisotropy, ~5 mT, is orders of magnitude smaller than the effective field of ~2 T, which can be estimated by the saturation field, raised by the RKKY interaction. These results, therefore, indicate that the remarkable change in the magnetization curves due to alternating stress shown in Fig. 2 originates from a much larger magnetic energy than that given by the stress-induced magnetic anisotropy.

Another possibility that explains the change in the magnetization curves with the alternating stress would be a biquadratic RKKY interaction [23,24] described by $-J_b\left(\mathbf{M}_j \cdot \mathbf{M}_{j+1}\right)^2$. While there are quite a few experimental observations and theoretical explanations for the biquadratic RKKY interaction, there have been no reports that stress would induce such interaction. Our present situation with the alternating stress does not fall into any of the theoretical models currently available. The small remanence induced in the sample with the alternating stress (See Figs. 2 (a) and (b)) indicating the emergence of the spontaneous magnetization, cannot be explained by the biquadratic RKKY with a positive $J_b$ as it does not stabilize any non-colinear spin configuration at zero field. Moreover, it is difficult for the biquadratic RKKY term to create a new magnetic order in superlattice systems regardless of positive or negative $J_b$ because the biquadratic term is a sub-leading magnetic interaction and $|J_b|$ is generally much smaller than the standard RKKY exchange coupling constant $J_{\mathrm{RKKY}}$. Therefore, it would rather be reasonable to consider it due to interlayer DM interaction induced by the symmetry breaking with the alternating stress.

Additionally, the stresses possibly induce some magnetic defects. Therefore, those defects can modify the domain structures and their dynamics to a certain extent, which could result in the modified magnetization curves presented here. But, if one compares the results with alternating stress (Fig. 2) and the uniform stress (Fig. 3b), it is noticeable that the non-zero remanence emerges with the alternating stress but no remanence emerges with the uniform stress. This suggests that the main root cause of the modified magnetization curves is not the stress itself but its spatial symmetry.

Since the spin orbit coupling that causes the DM interaction is strongly influenced by the electron correlation in the system [25], we need to pay attention to the symmetry of the local electric field, *e.g.* a distribution of the electron cloud at an atomic site, modified by the residual stress rather than the symmetry of the residual stress itself. We therefore presume that the direction of the modified local electric field $E_{\mathrm{loc}}$ is associated with the direction of the residual stress and analyze the possible emergence of the DM vectors by considering the symmetry of $E_{\mathrm{loc}}$ as discussed in detail in the supplementary materials [20]. Our symmetry analysis leads to the spatially staggered DM vectors perpendicular to the film normal in the magnetic superlattice with the alternating stress as depicted in Fig. 4.

To scrutinize the emergence of DM interaction and the staggered DM vectors in our superlattice, we conduct micromagnetic simulations based on Landau-Lifshits-Gilbert (LLG) equation incorporating the interlayer DM interaction energy (see Supplementary materials [20] for the detail formulations and parameters). We first reproduce the magnetization curve for $t_{\mathrm{Ru}}$ = 0.8 nm without stress with the parameters: $J_{\mathrm{RKKY}} = -1$ [mJ/m$^2$] and the saturation magnetization $M_s = 1000$ [emu/cc] as shown in Fig. 4 (a). We then put the uniform stress-induced uniaxial magnetic anisotropy energy of $K_s = 2.7 \times 10^3$[J/m$^3$] into the simulation and obtain almost identical magnetization curve to the one without the strains, which is in good agreement with the experimental results shown in Fig. 3 (b). It is also confirmed that the alternating uniaxial magnetic anisotropy presumably induced by the alternating stress does not change the magnetization curve at all (see the red curve in Fig. 4 (a)).

On the other hand, the magnetization curves with various interlayer DM interaction energy shown in Fig. 4 (b) reproduce well the experimentally observed characteristics, such as rapid increase of magnetization in the low field and the increase of the saturation field. We note that the simulation incorporates the staggered DM vectors shown in the inset of Fig. 4. By comparing these simulations with the experimental results, the interlayer DM interaction induced by the alternating strains is estimated to be as large as a few tens of percents of the RKKY interaction, that is $E_{\mathrm{IL-DM}} \sim 0.3$[mJ/m$^2$], which is an order of magnitude larger than the previously reported values [14,15]. While the spin orbit coupling in our system would not be drastically different compared with the previously reported systems using a similar heavy metal as a spacer layer, the large interlayer DM interaction can be attributed to the explicit symmetry breaking induced by the alternating stress.

The experimental observation and theoretical analyses therefore suggest that our magnetic superlattice with the alternating residual stress possibly host a staggered order of the DM vectors which has only been found in particular crystals before. Thanks to our peculiar stress application mechanism during the film depositions, spatially varied residual stress adds a completely new control on the microscopic structural symmetry in the magnetic superlattice. Since this interlayer DM interaction, or the DM vector, does not rely on the symmetry breaking at interfaces but the entire symmetry of the neighboring pair of the Co layers, it is expected to be scalable with the repetition of the layers.

In conclusion, we create Co/Ru magnetic superlattices with the residual alternating stress. Experimental observations and theoretical analyses indicate that the emergence of a strong interlayer DM interaction between Co layers through a Ru layer due to the symmetry breaking by the spatially varied residual stress. Moreover, from the viewpoint of the rotational symmetry and the translational symmetry in the system, we conclude that the DM vectors each of which resides between the neighboring pair of the Co layers should have the staggered order. Our results essentially indicate that the spatially varying strain of only 0.8% is quite effective to break the symmetry of the system and prove that the peculiar technique of our stress mechanism for the ultra-high vacuum deposition system perfectly works for artificially introducing the symmetry breaking. This technical breakthrough and the present results open a new paradigm of straintronics not only for magnetic superlattice systems but also for any kinds of different multilayered systems having order parameters which are affected by microscopic structural symmetries, such as semiconductors, ferroelectrics, and superconductors. The technique is, of course, highly compatible with flexible electronics, particularly for applications involving micro- and nano-patterned integrated circuitry on flexible substrates.

**Acknowledgements**

This work was supported in part by JST FOREST Program Grant Number JPMJFR2242, JST CREST Grant Number JPMJCR24R5, and JSPS KAKENHI Grants Nos. 23KK0093, 24H02235, 25K07198, 25H02112, and 26K21745. A part of the work was supported by "Advanced Research Infrastructure for Materials and Nanotechnology in Japan (ARIM)" of the Ministry of Education, Culture, Sports, Science and Technology (MEXT). Proposal Numbers JPMXP1225NU5208 and JPMXP1226NU5226.

**Figure captions**

**Figure 1: Schematic illustration of the film deposition technique with the alternating stress mechanism and the Co/Ru superlattice with the spatially alternating compressive stress.** (a) The alternating stress mechanism to obtain the alternating built-in strain in the layers. (b) The Co/Ru superlattice with the residual alternating stress. The thick white arrows depict the compressive residual stress in the films and the thin red arrows depict the spin $\mathbf{M}_j$ in the Co layers. The coordinate system is also depicted. (c) X-ray reflectivity measurements for the superlattice with $t_{\mathrm{Ru}}$ = 1.0, 0.8, and 0.5 nm.

**Figure 2: Comparison of magnetization curves for PEN sub./ Ru 2.0 nm/[Co 2.0 nm/ Ru $t_{\mathrm{Ru}}$ nm]$_9$/ Co 2.0 nm/ Ru 5.0 nm superlattice with and without the residual alternating stress.** Magnetization curves measured with a field along $x$ direction for the sample with $t_{\mathrm{Ru}}$ = 0.5 nm (a), 0.8 nm (c), and 1.0 nm (e), and their magnification of the first quadrant (b), (d), and (f), respectively.

**Figure 3: Effect of stress induced magnetic anisotropy on magnetization curves.** (a) Magnetization curves measured with the field along and perpendicular to the uniform strain for PEN sub./ Ru 2.0/Co 2.0 nm/Ru 2.0 nm. (b) Magnetization curves with the field along and perpendicular to the spatially uniform stress for PEN sub./ Ru 2.0 nm/ [Co 2.0 nm/ Ru $t_{\mathrm{Ru}}$ nm]$_9$/ Co 2.0 nm/ Ru 5.0 nm.

**Figure 4: Micromagnetic simulations for the [Ru/Co]$_{10}$ superlattice.** (a) Simulated magnetization curves with a spatially uniform strain induced magnetic anisotropy (green),

with the alternating strain induced magnetic anisotropy (red), and without any strain induced magnetic anisotropy (blue). (b) Simulated magnetization curves with the interlayer DM interaction of $|\mathbf{D}|/J_{RKKY} =$ 0, 0.1, 0.3, and 0.5. The illustration on the right depicts the staggered DM vectors in the superlattice incorporated in the simulation. The white arrows represent the magnetization and the black arrows represent the DM vector.

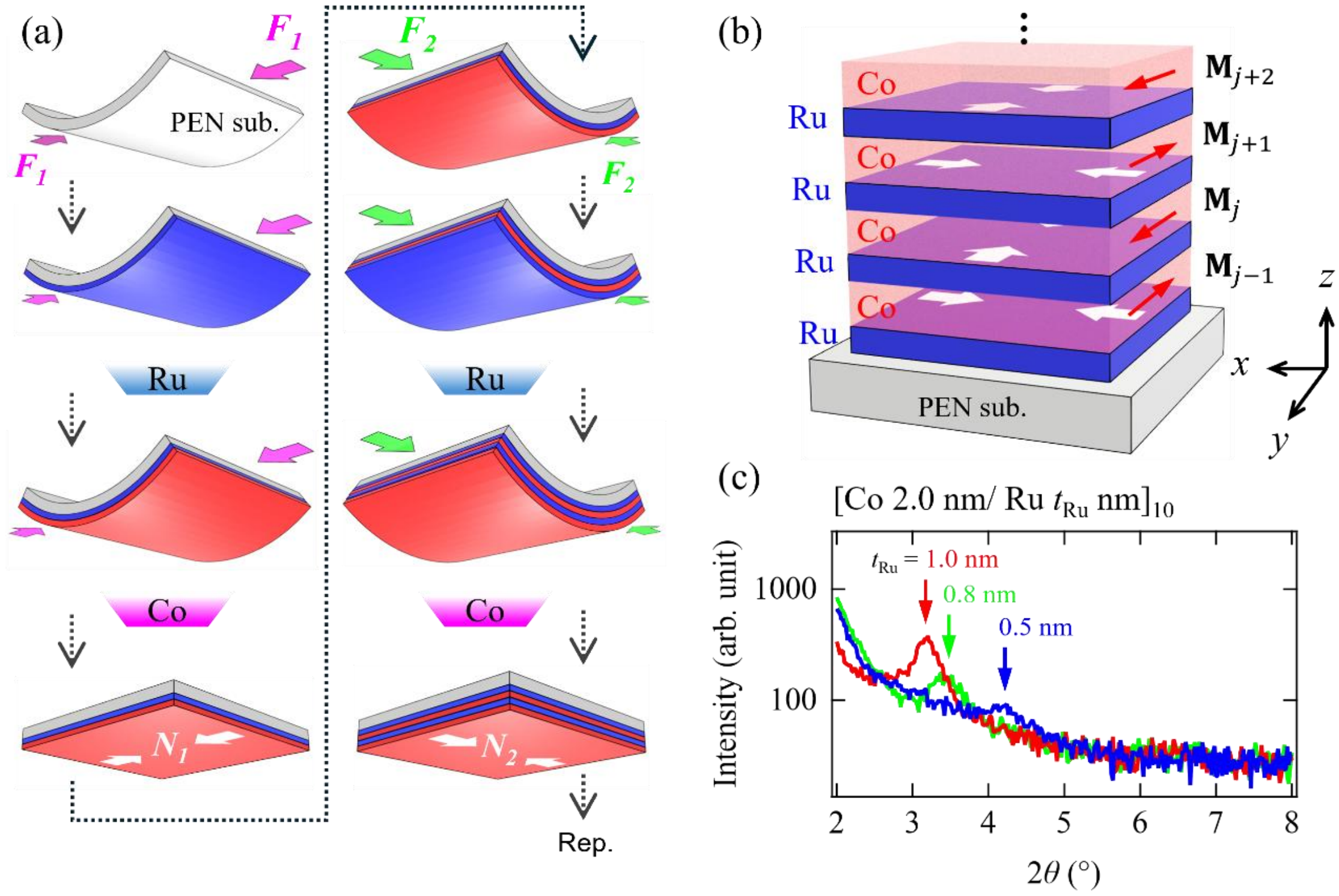

**Figure 1** Iwata et al.

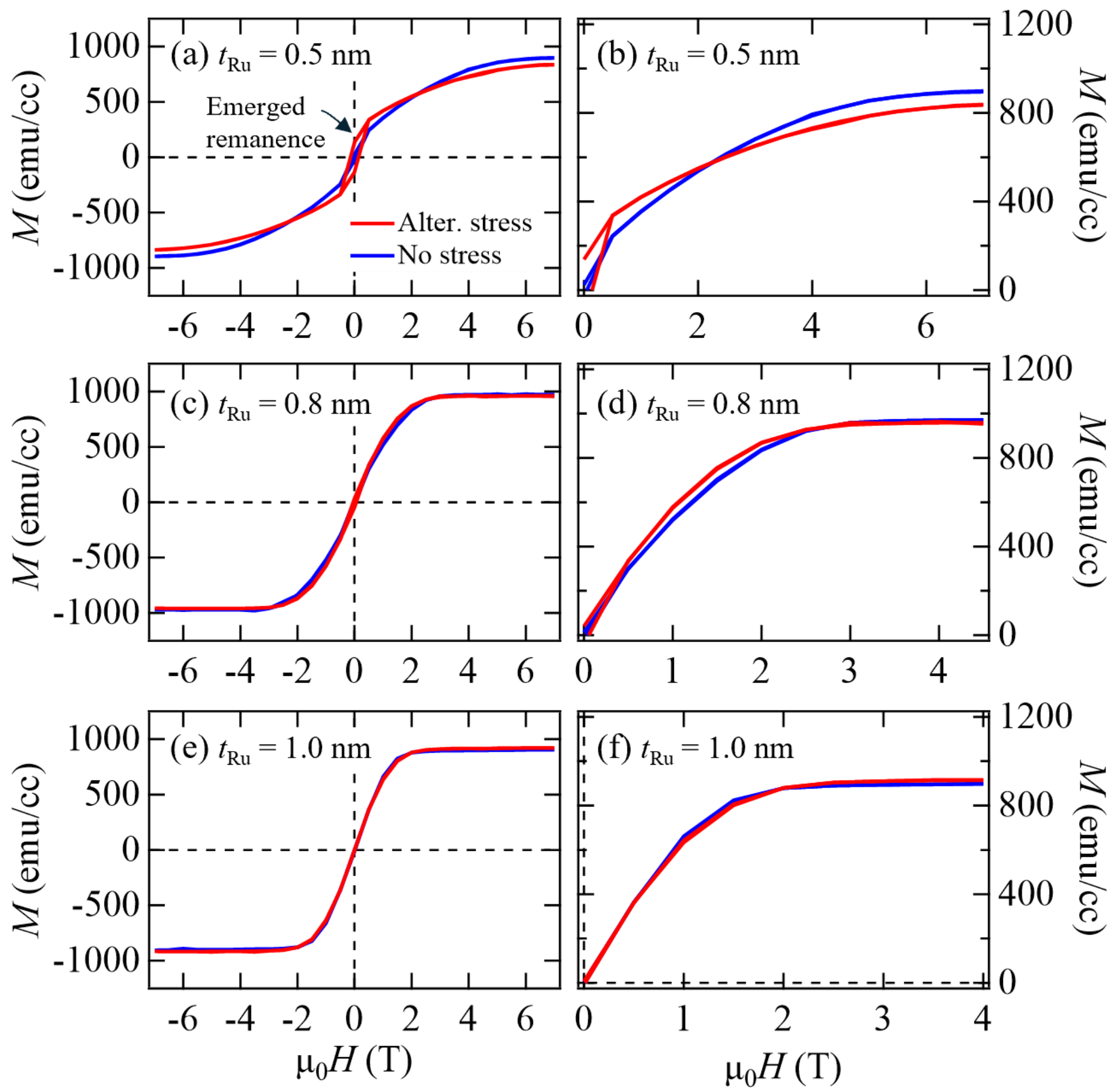

**Figure 2** Iwata et al.

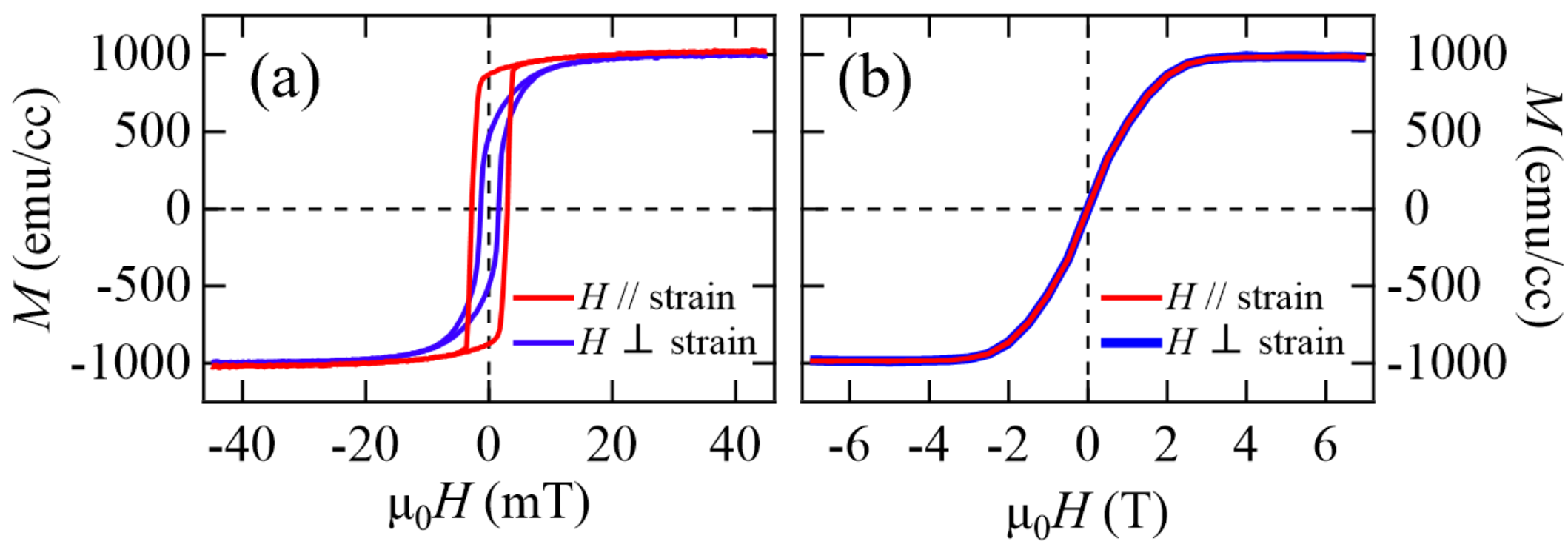

**Figure 3** Iwata et al.

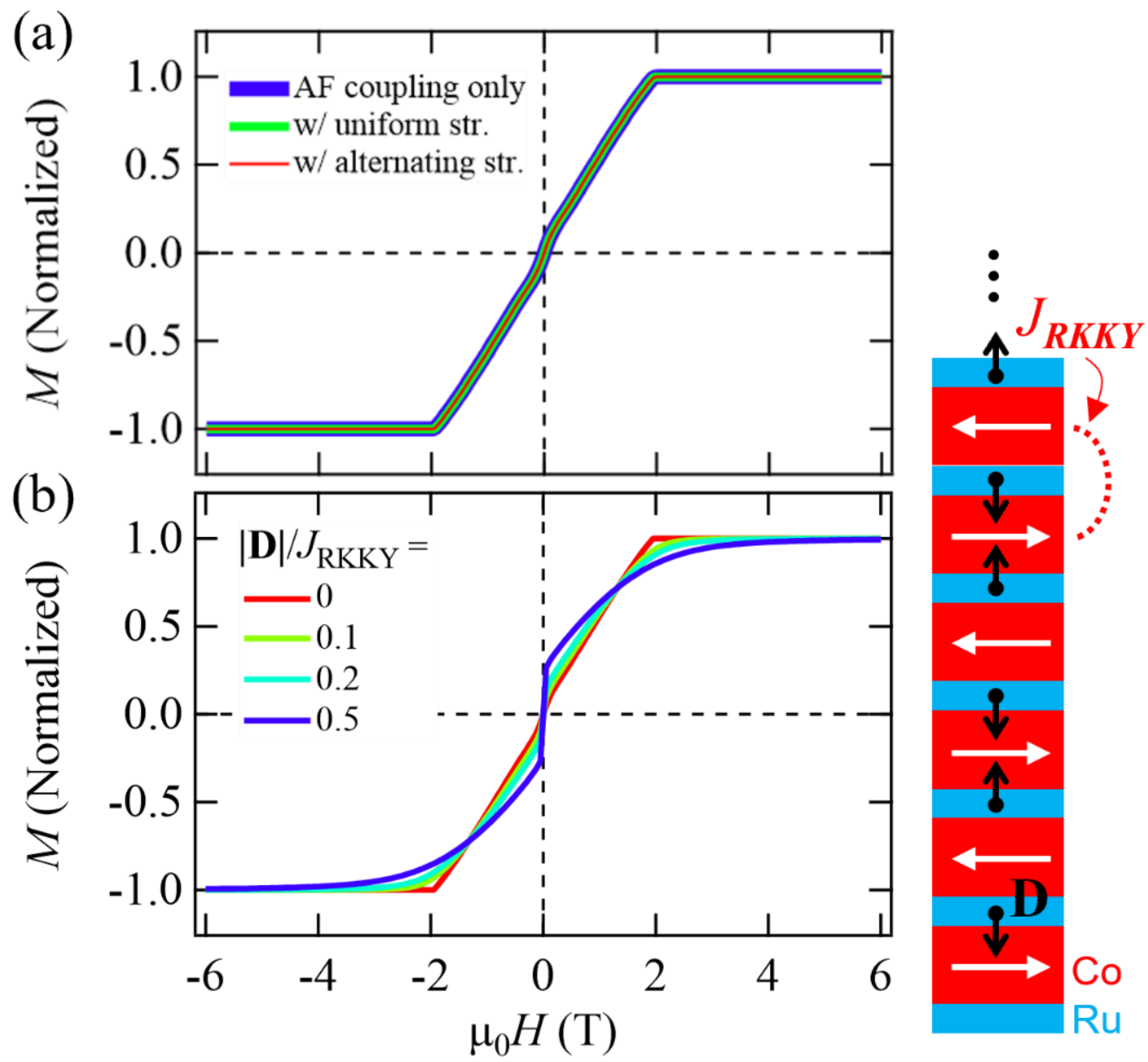

**Figure 4** Iwata et al.

# Giant staggered Dzyaloshinskii-Moriya vectors emerged in synthetic antiferromagnets with spatially alternating stress

Yuuga Iwata[1], Kensuke Hayashi[1], Reiju Furumaki[1], Satoshi Iihama[1], Masahiro Sato[2], and Takahiro Moriyama[1,3†]

[1]*Department of Materials Physics, Nagoya University, Nagoya, Aichi 464-8603, Japan*
[2]*Department of Physics, Chiba University, Chiba, Chiba 263-8522, Japan*
[3]*Research Center for Crystalline Materials Engineering, Nagoya University, Nagoya, Aichi 464-8603, Japan*

[†] Corresponding to moriyama.takahiro.a4@f.mail.nagoya-u.ac.jp

## Supplementary Materials

Index

## 1. Technical description of the strain mechanism

We employ a 0.1 mm-thick polyethylene naphthalate (PEN) sheet as a flexible substrate. The substrate was cut into 20 x 20 $mm^2$ square and set into a home-made stress mechanism designed to apply a compressive force that smoothly bends the substrate in two orthogonal directions. Figure 1 and Movies 1 and 2 (in the separate files) show the design and the actual motion of the substrate holder implemented with the strain mechanism. The special design enables the bending directions to change by the substrate rotation directions. The compressive force is applied by bumping and jamming the two legs sticking out of the holder. The direction of the substrate rotation, either clockwise or counterclockwise, is programmable by our sputtering system. Therefore, the controls and selections of the bending directions in each deposited layer is completely automatic. This mechanism not only creates the spatially alternating strains and uniform strains discussed in the main text, but also any designs of the strain directions in the whole superlattice.

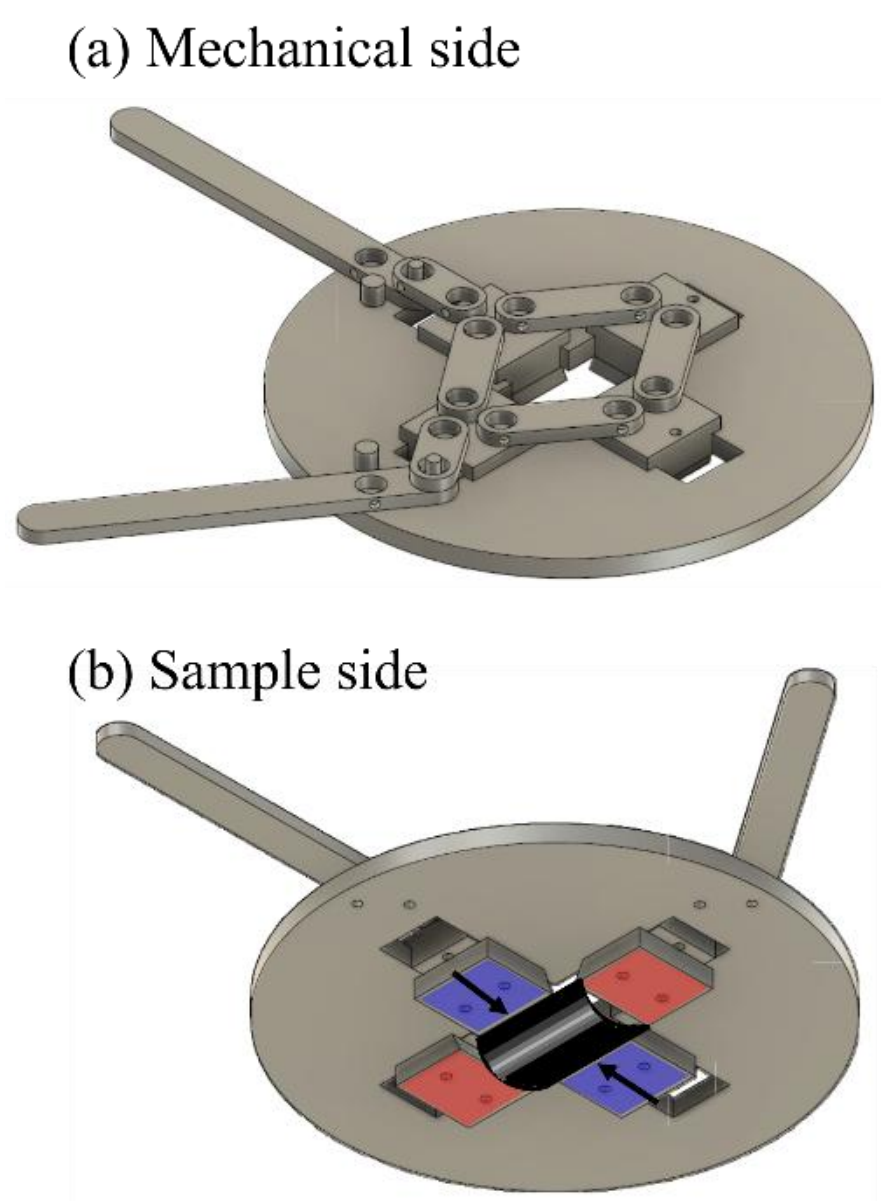

**Figure S1** The strain mechanism

## 2. Macroscopic models for Co/Ru superlattices

In this section, we construct the model describing macroscopic magnetism of the Co/Ru superlattices, based on symmetry and coarse graining arguments.

### A) Local electric field induced by the stress and the magnetic Hamiltonian

As shown in Fig. 1 of the main text, we introduce strains to Ru/Co superlattice systems in alternating or uniform manners (See also Fig. S2(a) and (b)). As a result, after the stress is introduced, the distributions of the internal local electric fields in both the Co and Ru layers are expected to differ from those in the same layers before the distortion. The schematic images of the local electric fields in both the alternating and uniform stress cases are shown in Fig. S2 (c) and (d). The arrows represent the direction of the local electric field $E_{\mathrm{loc}}$ at a local position of interest (*i.e.* atomic position) in each layer. Since the stress is induced parallel to the *x*-*y* plane, direction of $E_{\mathrm{loc}}$ is assumed to be parallel to the *x*-*y* plane. $E_{\mathrm{loc}}$ could be in any direction of the *x*-*y* plane at different local positions in each layer. The important point is that the directions of $E_{\mathrm{loc}}$ in neighboring Co (Ru) layers are perpendicular with each other in the alternating stress case, whereas they are parallel in the uniform stress case.

Under the above assumption, let us consider a simple model for macroscopic magnetism of Co layers. First, we suppose that the macroscopic magnetic moment in the $j$-th Co layer may be approximated by a single vector $\mathbf{M}_j$ because the Co crystal is a typical ferromagnetic metal. This is a sort of a coarse graining process from the renormalization-group (RG) viewpoint [1]. It is well known that the arguments based on RG or Ginzburg-Landau picture are powerful to understand the nature of macroscopic order parameters in many-body systems like magnets [2,3]. The macroscopic moments $\mathbf{M}_j$ are depicted in Fig. S2 and hereafter we merely call them "spins". Note that the index $j$ of spin $\mathbf{M}_j$ means the position of each Co layer and does not stand for the position of a certain Co atom in each layer. The magnetic interaction between neighboring spins, $\mathbf{M}_j$ and $\mathbf{M}_{j+1}$, is expected to emerge due to the conducting electron propagation in the Ru layer located between the $j$- and ($j$ + 1)-th Co layers. The dominant interaction would be the Heisenberg-type exchange interaction, $\mathbf{M}_j \cdot \mathbf{M}_{j+1}$. This may be referred to as a generalization of the RKKY interaction [2]. In addition to the exchange interaction, a Dzyaloshinskii-Moriya (DM) interaction [3,4] is expected to appear between neighboring spins. The stress-induced electric field generally reduces the symmetry of the superlattice resulting in a broken inversion symmetry by $E_{\mathrm{loc}}$. In such a lower-symmetric magnetic system, the DM interaction is often allowed to appear. In fact, a recent theoretical study [5] shows that the DM interaction can emerge by applying an intense electric field to a broad class of Mott insulators because of the electric-field induced spin-orbit (SO) interaction. From these arguments, the effective Hamiltonian describing macroscopic magnetic moments $\mathbf{M}_j$ in the superlattices is given by,

$$\hat{H}_{\mathrm{Co-Ru}} = \hat{H}_{\mathrm{ex}} + \hat{H}_{\mathrm{DM}} = \sum_j J_{j,j+1}\, \mathbf{M}_j \cdot \mathbf{M}_{j+1} + \sum_j \mathbf{D}_{j,j+1} \cdot \left(\mathbf{M}_j \times \mathbf{M}_{j+1}\right), \tag{S1}$$

where $J_{j,j+1}$ is the strength of the exchange interaction $\hat{H}_{\mathrm{ex}}$ and $\mathbf{D}_{j,j+1}$ is the DM vector in the DM interaction $\hat{H}_{\mathrm{DM}}$. The values of $J_{j,j+1}$ and $\mathbf{D}_{j,j+1}$ generally depend on the manner of stress introduction. In the next subsections, we try to restrict the $j$ (spatial) dependence of $J_{j,j+1}$ and $\mathbf{D}_{j,j+1}$, by using symmetries of the superlattice systems.

### B) Translational symmetry

First, we consider the translation symmetry along the $z$ axis. Namely, we impose a condition that the Hamiltonian is invariant under two-site translation $\mathbf{M}_j \to \mathbf{M}_{j+2}$ in the alternating-stress system, while it does under the one-site translation $\mathbf{M}_j \to \mathbf{M}_{j+1}$ in the uniform-stress system. As a result, we find the relation,

$$\begin{aligned} &J_{j,j+1} = J_{j+2,j+3},\ \mathbf{D}_{j,j+1} = \mathbf{D}_{j+2,j+3},\ \text{(alternating stress)} \\ &J_{j,j+1} = J_{j+1,j+2},\ \mathbf{D}_{j,j+1} = \mathbf{D}_{j+1,j+2}.\ \text{(uniform stress)} \end{aligned} \tag{S2}$$

Here, we note that one site means one layer. Strictly speaking, the translational symmetry is broken if the system length along $z$ axis is finite. However, if the superlattice structure is sufficiently long and we focus on only bulk quantities (not physical nature near the edge), the conclusion derived from the translation symmetry would be almost correct. In fact, the translation symmetry has been often utilized in studies of other artificial systems such as semiconductor superlattices and Moiré systems.

### C) Superlattice with the spatially alternating stress

In this subsection, we focus on the symmetries in the superlattice with the spatially alternating stress, which is the main target of the present work. We consider the combination of three symmetry operations depicted in Fig. S3. One finds from the figure that the stress-induced electric-field distribution is invariant under the combination. Therefore, the Hamiltonian $\hat{H}_{\mathrm{Co-Ru}}$ should also be invariant under the same operations.

In the following, we discuss how the symmetry gives constraint to the $j$ dependence of $J_{j,j+1}$ and $\mathbf{D}_{j,j+1}$. To this end, it is convenient to consider the local Hamiltonians

$$\widehat{H}_{j,j+1}^{\mathrm{ex}} = J_{j,j+1}\mathbf{M}_j \cdot \mathbf{M}_{j+1}, \tag{S3}$$

$$\widehat{H}_{j,j+1}^{\mathrm{DM}} = \mathbf{D}_{j,j+1} \cdot \left(\mathbf{M}_j \times \mathbf{M}_{j+1}\right), \tag{S4}$$

where the relations $\widehat{H}_{\mathrm{ex}} = \sum_j \widehat{H}_{j,j+1}^{\mathrm{ex}}$ and $\widehat{H}_{\mathrm{DM}} = \sum_j \widehat{H}_{j,j+1}^{\mathrm{DM}}$ hold.

First, we perform the π/2 rotation around the $z$ axis [Fig. S3 (b)] for the system. It leads to transform $M^x \rightarrow M^y$ and $M^y \rightarrow -M^x$. Therefore, the local Hamiltonians are transformed as

$$\widehat{H}_{j,j+1}^{\mathrm{ex}} \rightarrow \widehat{H}_{j,j+1}^{\mathrm{ex}} =: \widehat{H}_{j,j+1}^{\mathrm{ex(b)}}, \tag{S5}$$

$$\begin{aligned}\widehat{H}_{j,j+1}^{\mathrm{DM}} \rightarrow & -D_{j,j+1}^{x}\left(M_j^x M_{j+1}^z - M_j^z M_{j+1}^x\right) \\ & + D_{j,j+1}^{y}\left(M_j^z M_{j+1}^y - M_j^y M_{j+1}^z\right) - D_{j,j+1}^{z}\left(M_j^y M_{j+1}^x - M_j^x M_{j+1}^y\right) \\ & =: \widehat{H}_{j,j+1}^{\mathrm{DM(b)}}.\end{aligned} \tag{S6}$$

The exchange interaction is unchanged. Secondly, we apply the π rotation around the $x$ axis [Fig. S3 (c)], in which spins are changed as $M^{y,z} \rightarrow -M^{y,z}$ and the site indices of spins are transformed as $j-n \leftrightarrow j+1+n$ ($n$ is an arbitrary integer). As a result, the local Hamiltonians are mapped as follows:

$$\widehat{H}_{j,j+1}^{\mathrm{ex(b)}} \rightarrow \widehat{H}_{j+1,j}^{\mathrm{ex}} = \widehat{H}_{j,j+1}^{\mathrm{ex}} =: \widehat{H}_{j,j+1}^{\mathrm{ex(c)}}, \tag{S7}$$

$$\begin{aligned}\widehat{H}_{j,j+1}^{\mathrm{DM(b)}} \rightarrow & D_{j,j+1}^{x}\left(M_{j+1}^x M_j^z - M_{j+1}^z M_j^x\right) \\ & + D_{j,j+1}^{y}\left(M_{j+1}^z M_j^y - M_{j+1}^y M_j^z\right) + D_{j,j+1}^{z}\left(M_{j+1}^y M_j^x - M_{j+1}^x M_j^y\right) \\ & =: \widehat{H}_{j,j+1}^{\mathrm{DM(c)}}.\end{aligned} \tag{S8}$$

Note that the exchange interaction is again invariant under the operation (c). Third, we apply the $x$-$y$ plane mirror reflection with respect to the Co layer with the moment $\mathbf{M}_{j+1}$ [see Fig. S3 (c) and (d)]. By comparing Fig. S3 (a) and (d), we see that the system (the electric-field distribution) returns to the original system after the operation (d). The final reflection induces $M^{x,y} \rightarrow -M^{x,y}$ and $j-n \leftrightarrow j+2+n$ because the spin is an axial vector. Thus, the local Hamiltonians are transformed as

$$\widehat{H}_{j,j+1}^{\mathrm{ex(c)}} \rightarrow J_{j,j+1}\mathbf{M}_{j+2} \cdot \mathbf{M}_{j+1} =: \widehat{H}_{j,j+1}^{\mathrm{ex(d)}}, \tag{S9}$$

$$\begin{aligned}\widehat{H}_{j,j+1}^{\mathrm{DM(c)}} \rightarrow & D_{j,j+1}^{y}\left(M_{j+1}^y M_{j+2}^z - M_{j+1}^z M_{j+2}^y\right) \\ & + D_{j,j+1}^{x}\left(M_{j+1}^z M_{j+2}^x - M_{j+1}^x M_{j+2}^z\right) - D_{j,j+1}^{z}\left(M_{j+1}^x M_{j+2}^y \right. \\ & \left. - M_{j+1}^y M_{j+2}^x\right). =: \widehat{H}_{j,j+1}^{\mathrm{DM(d)}}.\end{aligned} \tag{S10}$$

The transformed Hamiltonian must be equivalent to the original one after the three operations (b)-(d). To perform the quantitative comparison between these two Hamiltonians, we should shift all site indices in $\widehat{H}_{j,j+1}^{\mathrm{ex(d)}}$ and $\widehat{H}_{j,j+1}^{\mathrm{DM(d)}}$ as $j \rightarrow j-1$ [compare Fig. S3 (a) and (d)], which is merely equivalent to renaming operators and coupling constants. We define the shifted local Hamiltonians as $\widehat{H}_{j,j+1}^{\mathrm{ex(e)}} \coloneqq \widehat{H}_{j-1,j}^{\mathrm{ex(d)}}$ and $\widehat{H}_{j,j+1}^{\mathrm{DM(e)}} \coloneqq \widehat{H}_{j-1,j}^{\mathrm{DM(d)}}$. Note that this "renaming" operation includes translating the site index of the coupling constants such as $J_{j,j+1} \to J_{j-1,j}$ and $\mathbf{D}_{j,j+1} \to \mathbf{D}_{j-1,j}$.

After the renaming, the local Hamiltonians should satisfy the following equations:

$$\hat{H}^{\mathrm{ex(e)}}_{j,j+1} = \hat{H}^{\mathrm{ex}}_{j,j+1}\,, \hat{H}^{\mathrm{DM(e)}}_{j,j+1} = \hat{H}^{\mathrm{DM}}_{j,j+1}. \tag{S11}$$

From these relations, we obtain

$$J_{j,j+1} = J_{j-1,j}. \tag{S12}$$

$$D^x_{j,j+1} = D^y_{j-1,j},\ D^y_{j,j+1} = D^x_{j-1,j},\ D^z_{j,j+1} = -D^z_{j-1,j}. \tag{S13}$$

Combining these with Eq. S2, we find the following equations

$$J_{j,j+1} = J. \tag{S14}$$

$$D^x_{j,j+1} = D^y_{j-1,j} =: D^{(\mathrm{a})}_j\ ,\ \ D^y_{j,j+1} = D^x_{j-1,j} =: D^{(b)}_j, \tag{S15}$$

$$D^z_{j,j+1} = (-1)^j D^z. \tag{S16}$$

That is, the exchange interaction is spatially uniform, while the z component of the DM vector is staggered [6,7]. Since the macroscopic spins $\mathbf{S}_j$ in the superlattice tend to be in the *x*-*y* plane due to a strong demagnetizing field, the *x* and *y* components of the DM vector, $D^{x,y}$, do not effectively contribute to the energy of the DM interaction. From these symmetry arguments, we find that the relevant form of the Hamiltonian in the superlattice with the alternating stress is given by

$$\hat{H}_{\mathrm{Co-Ru}} = \sum_j J\,\mathbf{M}_j \cdot \mathbf{M}_{j+1} + \sum_j (-1)^j D^z \left(\mathbf{M}_j \times \mathbf{M}_{j+1}\right)^z, \tag{S17}$$

Finally, we comment on the reflection with respect to the *z*-*x* (or *y*-*z*) plane. It seems to be possible to construct the symmetry operation including these reflections, under which the system or electric-field distribution is invariant. However, the relative relationship between the electric fields at two arbitrary points in each layer may change by the reflections. Therefore, we guess that the symmetry operation including these reflections is too strong to restrict the forms of exchange and DM interaction. In fact, we can lead to $\mathbf{D}_{j,j+1} = 0$ by combining the *z*-*x* plane reflection with other symmetries. As we will show in the following section, the absence of DM interaction is inconsistent with the experimental observation of the magnetization process. Thus, we conclude that it is reasonable to consider only the translation, the rotations and the mirror reflection with respect to the *x*-*y* plane when we restrict the magnetic interactions.

**D) Superlattice with the spatially uniform stress**

Similarly to the former subsection, we consider the symmetry operation that makes the superlattice with the uniform stress invariant. In this system, one easily finds that the system is invariant under the mirror reflection with respect to the *x*-*y* plane, as shown in Fig. S4. Through this reflection, spins are transformed to $M^{x,y} \to -M^{x,y}$ and the site indices of spins are changed to $j - n \to j + 1 + n$ because the spin is an axial vector. The local Hamiltonians are therefore transformed as

$$\hat{H}^{\mathrm{ex}}_{j,j+1} \to \hat{H}^{\mathrm{ex}}_{j,j+1} =: \hat{H}^{\mathrm{ex(b)}}_{j,j+1}, \tag{S18}$$

$$\hat{H}^{\mathrm{DM}}_{j,j+1} \to -D^{x}_{j,j+1}\left(M^{y}_{j}M^{z}_{j+1} - M^{z}_{j}M^{y}_{j+1}\right) - D^{y}_{j,j+1}\left(M^{z}_{j}M^{x}_{j+1} - M^{x}_{j}M^{z}_{j+1}\right) + D^{z}_{j,j+1}\left(M^{x}_{j}M^{y}_{j+1} - M^{y}_{j}M^{x}_{j+1}\right) =: \hat{H}^{\mathrm{DM(b)}}_{j,j+1}. \tag{S19}$$

The equivalence between the original and transformed Hamiltonians imposes the relation

$$D^{z}_{j,j+1} = 0, \tag{S16}$$

while it does not require any constraint for the $x$ and $y$ components of the uniform DM vector. Similarly to the case of the alternating stress, $D^{x,y}$ is irrelevant in the energetic sense because of the large demagnetizing field. Therefore, in the case of the uniform stress, the relevant Hamiltonian is a simple Heisenberg model,

$$\hat{H}_{\mathrm{Co-Ru}} = \sum_{j} J\,\mathbf{M}_j \cdot \mathbf{M}_{j+1}. \tag{S17}$$

## 3. Micromagnetic simulations with staggered D vectors

Magnetization curves of multilayers were simulated as discussed in the main text. The equilibrium magnetization configurations were obtained by calculating Landau-Lifshitz-Gilbert equation as follows,

$$\frac{\mathrm{d}\mathbf{M}_j}{\mathrm{d}t} = -\gamma\mu_0\mathbf{M}_j \times \mathbf{H}_{\mathrm{eff},j} + \alpha\mathbf{M}_j \times \frac{\mathrm{d}\mathbf{M}_j}{\mathrm{d}t}, \tag{S18}$$

where $\mathbf{M}_j, \mathbf{H}_{\mathrm{eff},j}, \gamma, \alpha$ are unit magnetization vector at layer $j$, effective magnetic field at layer $j$, gyromagnetic ratio and damping parameter, respectively. Here, demagnetizing field at each layer, in-plane uniaxial magnetic anisotropy due to the stress, interlayer RKKY exchange interaction, and interlayer DM interaction with the staggered DM vector were considered in $\mathbf{H}_{\mathrm{eff},j}$ as,

$$\mathbf{H}_{\mathrm{eff},j} = H\hat{\mathbf{x}} - M_{\mathrm{s}}M^{z}_{j}\hat{\mathbf{z}} + H^{x}_{k,j}M^{x}_{j}\hat{\mathbf{x}} + H^{y}_{k,j}M^{y}_{j}\hat{\mathbf{y}} + \begin{cases} H_{\mathrm{ex},j+1}\mathbf{M}_{j+1} + H^{z}_{\mathrm{DM},j+1}\left(M^{y}_{j+1}\hat{\mathbf{x}} - M^{x}_{j+1}\hat{\mathbf{y}}\right) & (j=1) \\ H_{\mathrm{ex},j}\mathbf{M}_{j-1} - H^{z}_{\mathrm{DM},j}\left(M^{y}_{j-1}\hat{\mathbf{x}} - M^{x}_{j-1}\hat{\mathbf{y}}\right) & (j=n) \\ H_{\mathrm{ex},j}\mathbf{M}_{j-1} - H^{z}_{\mathrm{DM},j}\left(M^{y}_{j-1}\hat{\mathbf{x}} - M^{x}_{j-1}\hat{\mathbf{y}}\right) + H_{\mathrm{ex},j+1}\mathbf{M}_{j+1} + H^{z}_{\mathrm{DM},j+1}\left(M^{y}_{j+1}\hat{\mathbf{x}} - M^{x}_{j+1}\hat{\mathbf{y}}\right) & (j \geq 2\ \&\ j \leq n-1) \end{cases}, \tag{S19}$$

$H_{\mathrm{ex},j}$, $H^{z}_{\mathrm{DM},j}$, and $H_{k,j}$ are defined as $H_{\mathrm{ex},j} = J_{\mathrm{RKKY}}/(\mu_0 M_{\mathrm{s}} t_{\mathrm{Co}})$, $H^{z}_{\mathrm{DM},j} = D^{z}_{j,j+1}/(\mu_0 M_{\mathrm{s}} t_{\mathrm{Co}})$, and $H_{k,j} = 2K_{\mathrm{s},j}/\mu_0 M_{\mathrm{s}}$, where $t_{\mathrm{Co}}$ and $M_{\mathrm{s}}$ are the thickness and the saturation magnetization of the Co layer, respectively. $H$ is a strength of in-plane external magnetic field. $n$ is the total number of the magnetic layers. Here, as mentioned in the above section, staggered DM interaction is considered as $D^{z}_{j,j+1} = (-1)^{j}D^{z}$. Also, uniaxial magnetic anisotropy due to staggered stress is considered as,

$$H^{x}_{k,j} = \begin{cases} 2K_{\mathrm{s}}/M_{\mathrm{s}} & (\mathrm{mod}(j,2)=0) \\ 0 & (\mathrm{mod}(j,2)=1) \end{cases}, \quad H^{y}_{k,j} = \begin{cases} 0 & (\mathrm{mod}(j,2)=0) \\ 2K_{\mathrm{s}}/M_{\mathrm{s}} & (\mathrm{mod}(j,2)=1) \end{cases}. \tag{S20}$$

Time-integration of Eq. S18 was calculated by 4th order Runge-Kutta method until magnetization was converged. Parameters used in the calculation shown in Fig. 4 were $M_{\mathrm{s}} = 1000$ [emu/cc], $J_{\mathrm{RKKY}} = -1.0 \times 10^{-3}$ [J/m$^2$], $D^z = 0.1, 0.2, 0.5 \times 10^{-3}$ [J/m$^2$], and $K_{\mathrm{s}} = 2.7 \times 10^3$ [J/m$^3$].

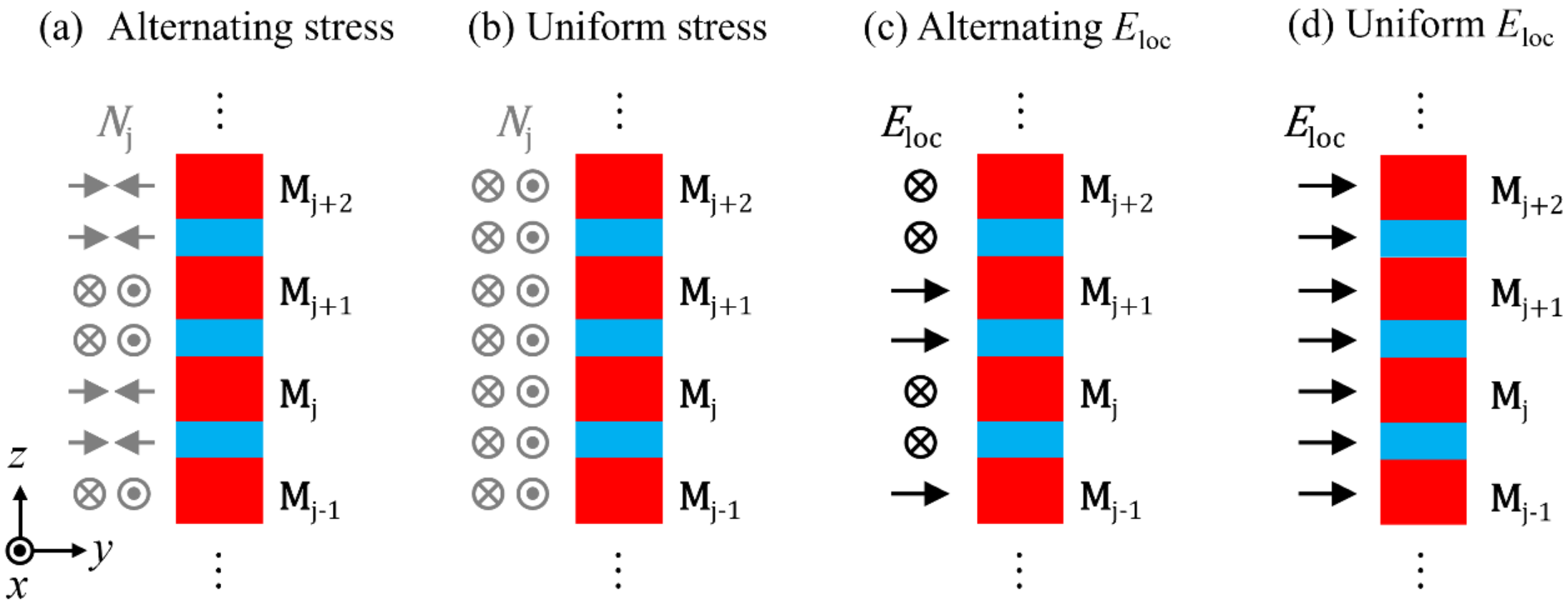

**Figure S2**

Magnetic superlattice with the alternating (a) and the uniform (b) stress. The gray arrows indicate the direction of the stress introduced in each layer. The vector $S_j$ represents the macroscopic magnetic moment on the *j*-th Co layer. The black arrows are the expected stress-induced local electric field $E_{loc}$ in the Co and Ru layers.

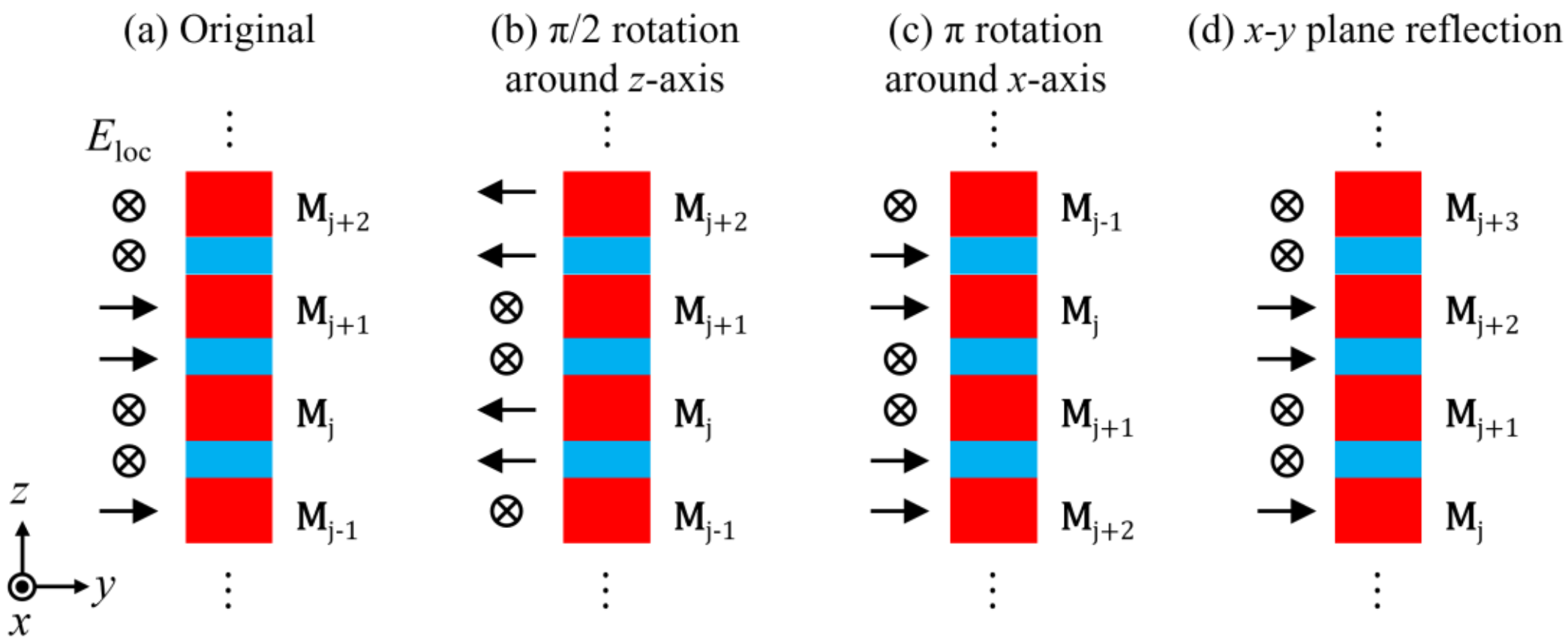

**Figure S3**

Combination of symmetry operations in the superlattice with the spatially alternating stress: (a) is the original electric-field and spin distributions. (b) shows the distributions after π/2 rotation around the *z* axis. (c) represents those after acting π/2 rotation around the *x* axis on the state (b). (d) is those after applying the *x*-*y* plane mirror reflection to the state (c).

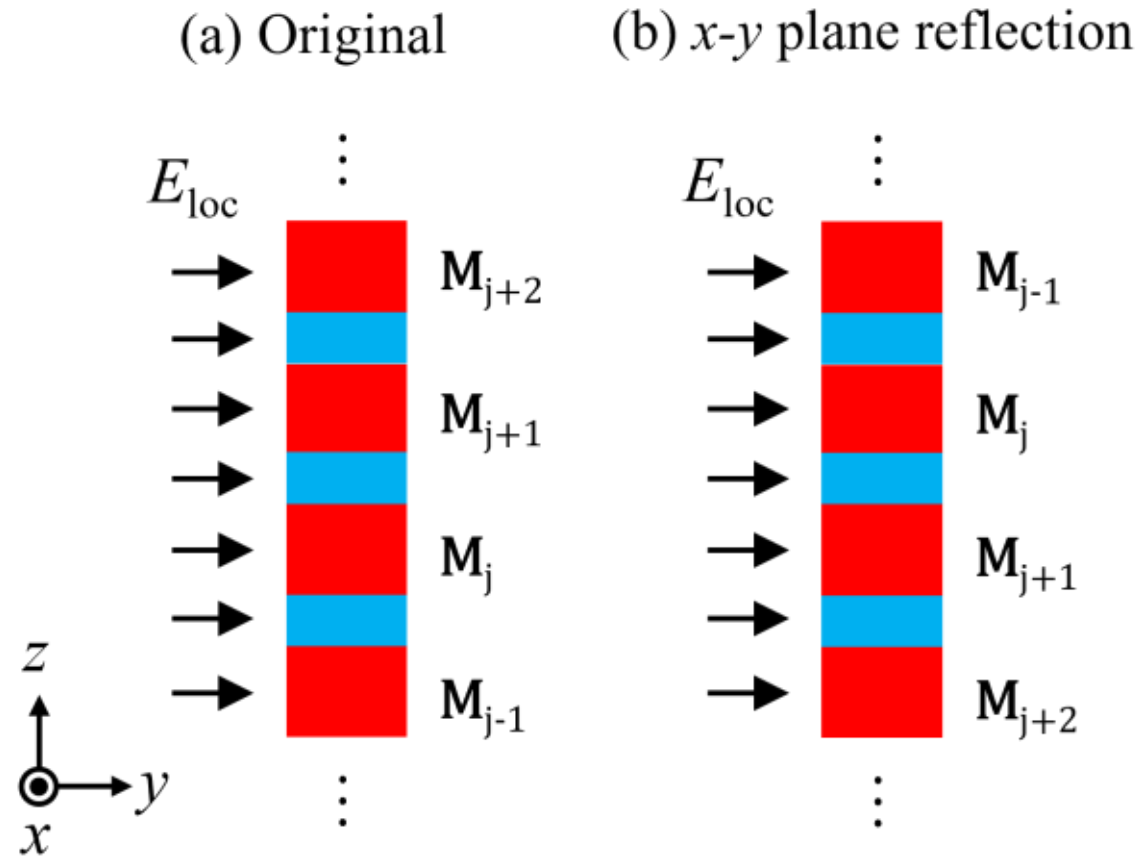

**Figure S4**

Reflection symmetry operation in the superlattice with the spatially uniform strains. (a) shows the original electric-field and spin distributions and (b) shows those after the *x*-*y* plane reflection.